\documentclass[twocolumn,showpacs,preprintnumbers,superscriptaddress,amsmath,amssymb,prb]{revtex4}

\usepackage{epsfig}    
\usepackage{dcolumn}
\usepackage{bm}

\def\etal{{\em et al.\/}}
\def\kp{${\bf k}\cdot{\bf p}$}
\def\eva3{$eV$$\cdot$\AA$^3$}               
\def\*#1*/{}                        

\def\deg{\nobreak\hbox{\hskip 2.0 truept\hbox{$^\circ$}}}
\def\ie{{\em i.e.\/}}

\def\oneh{\tfrac{1}{2}\tfrac{1}{2}0}
\def\oneq{\tfrac{1}{4}\tfrac{1}{4}\tfrac{1}{4}}
\def\moneq{\tfrac{-1}{4}\tfrac{-1}{4}\tfrac{-1}{4}}

\newcommand{\ket}[1]{|#1\rangle}
\newcommand{\braket}[2]{\langle #1|#2\rangle}
\newcommand{\melem}[3]{\left\langle #1 \left| #2 \right| #3 \right\rangle }
\newcommand{\refeq}[1]{Eq.~(\ref{#1})}

\begin{document}

\title{Higher order contributions to Rashba and Dresselhaus effects}

\author{X. Cartoix\`a}
\email[Electronic address: ]{Xavier.Cartoixa@uab.es}
\affiliation{
Computational Research Division, Lawrence Berkeley National Laboratory, Berkeley, CA 94720, USA
}%
\affiliation{
Departament d'Enginyeria Electr\`onica, Universitat Aut\`onoma de Barcelona, 08193 Bellaterra, Barcelona, Spain
}%

\author{L.-W. Wang}
\affiliation{
Computational Research Division, Lawrence Berkeley National Laboratory, Berkeley, CA 94720, USA
}%

\author{D. Z.-Y. Ting}%
\affiliation{
Jet Propulsion Laboratory, California Institute of Technology, Pasadena, CA 91109, USA
}%

\author{Y.-C. Chang}
\affiliation{
Department of Physics, University of Illinois at Urbana-Champaign, Urbana, IL 61801, USA
}%

\date{\today}

\begin{abstract}
We have developed a method to systematically compute the form of Rashba- and Dresselhaus-like contributions to the spin Hamiltonian of heterostructures to an arbitrary order in the wavevector {\bf k}. This is achieved by using the double group representations to construct general symmetry-allowed Hamiltonians with full spin-orbit effects within the tight-binding formalism. We have computed full-zone spin Hamiltonians for [001]-, [110]- and [111]-grown zinc blende heterostructures ($D_{2d},C_{4v},C_{2v},C_{3v}$ point group symmetries), which are commonly used in spintronics. After an expansion of the Hamiltonian up to third order in {\bf k}, we are able to obtain additional terms not found previously. The present method also provides the matrix elements for bulk zinc blendes ($T_d$) in the anion/cation and effective bond orbital model (EBOM) basis sets with full spin-orbit effects.
\end{abstract}

\maketitle




\section{Introduction}

There have been strong recent interests in spin effects in semiconductors for applications in spintronics~\cite{Prinz1995,WolfAwschalomBuhrman2001}. Particularly, spin splittings due to bulk inversion asymmetry~\cite{Dresselhaus1955} (BIA)---arising from the different chemical character of the anion and the cation in the zinc blende structure---and structural inversion asymmetry~\cite{BychkovRashba1984:physc} (SIA)---appearing in layered structures where ``top'' is different from ``down'': asymmetric composition, doping, etc.---, are studied because a good understanding of the intraband spin splittings is essential in any attempt to understand the operation of spintronic devices at a microscopic level, since they determine the spin dynamics during relaxation and transport processes~\cite{DyakonovPerel1971b,DyakonovKachorovskii1986,%
AverkievGolub1999,VoskoboynikovLinLee2000:jap,TingCartoixa2002,KogaNittaTakayanagi2002,%
CartoixaTingChang2003,SchliemannEguesLoss2003,CartoixaTingChang2005}. The pioneering work by Dresselhaus~\cite{Dresselhaus1955}, Bychkov and Rashba~\cite{BychkovRashba1984:physc}, and D'yakonov and Kachorovski\u{\i}\cite{DyakonovKachorovskii1986} taught us the functional form of the leading order contributions---up to first (third) order in the wavevector {\bf k} for SIA (BIA)---to the spin Hamiltonians in bulk and heterostructure zinc blendes due to the various sources of inversion asymmetry.

This constructive procedure is of great use as it describes the main physics but, as shown below, $\mathcal{O}(k^3)$ contributions for heterostructures due to BIA~\cite{DyakonovKachorovskii1986} should contain additional terms, while the $\mathcal{O}(k^3)$ contributions arising from SIA have not been studied. Thus, it is also important to have a systematic modeling tool which guarantees that all spin-related qualitative features in the band structures will be present. The empirical tight-binding method as formulated by Slater and Koster~\cite{SlaterKoster1954} provides such a systematic way of generating all the symmetry-allowed terms that can appear in a Hamiltonian. It has been used extensively in the computation of bulk, heterostructure and surface properties, and today finds widespread use in the study of nanostructures involving thousands and even millions of atoms~\cite{OyafusoKlimeckBowen2003}, or transport properties~\cite{HamadaSawadaOshiyama1992}.

However, the original formulation by Slater and Koster was obtained through the use of single group symmetry operations, which prevented it from describing spin effects. Later, Chadi~\cite{Chadi1977} extended the original method with an on-site spin-orbit energy which effectively describes the zone center split-off splitting and breaks the double degeneracy of the bands for systems without an inversion center. This scheme contains the essential features of spin-orbit interaction, but does not reproduce some qualitative aspects such as the linear spin splitting in valence bands in zinc blendes~\cite{Dresselhaus1955}. Boykin~\cite{Boykin1998} has introduced supplementary matrix elements between nearest-neighbor atomic orbitals as a remedy for zinc blende structures, but that procedure does not guarantee {\it a priori} that all spin-orbit effects will be considered, and does not treat systematically other types of structures. The development presented in this paper allows us to address these issues precisely.

In this Article, we compute the forms of SIA and BIA contributions to the spin Hamiltonian of common heterostructures, beyond the conventional Rashba~\cite{BychkovRashba1984:physc} and Dresselhaus~\cite{Dresselhaus1955,DyakonovKachorovskii1986} terms, to an arbitrary order in the wavevector {\bf k}. This allows us to treat SIA on the same footing as BIA, providing higher order corrections to the Rashba Hamiltonian for [001] ($D_{2d},C_{4v},C_{2v}$), [110] ($C_{2v}, C_s$) and [111] ($C_{3v}$) diamond and zinc blende quantum wells. We achieve this by following the constructive process of Slater and Koster~\cite{SlaterKoster1954} (SK), but using double group irreducible representations (irreps) as opposed to single group irreps. We also apply the method to the generation of bulk zinc blende ($T_d$) Hamiltonians in the anion/cation and effective bond orbital model (EBOM)~\cite{Chang1988,EinevollChang1989} basis sets containing full spin-orbit effects. The procedure presented could, of course, also be used for the description of the bands of materials involving heavy elements (i.e. large relativistic effects), such as lead compounds, rare earths, etc.


\section{Methods}
\label{sec:methods}

In order to achieve our goal of constructing spin Hamiltonians to arbitrary order in the wavevector {\bf k}, we first construct the tight binding Hamiltonian for the corresponding point group symmetry using the EBOM basis, and then series-expand in {\bf k} to the desired order. Also, since we are interested in the correct description of relativistic and, in particular, spin effects in the electronic bands, we are led naturally to the use of double group irreps~\cite{Falicov1966} to describe the symmetry operations of the crystal. Thus, if we follow the SK procedure~\cite{SlaterKoster1954} using double as opposed to single groups, we will be assured to obtain the most general Hamiltonian compatible with the crystal symmetries which, by construction, will include all spin-orbit effects.

The basis states of double group irreps are much less amenable to brute force operation than their single group counterparts. Therefore, it becomes necessary to write the tight binding equations and the symmetry constraints of the parameters in a form that allows computer manipulation. This is done in Appendices~\ref{ssec:ETB_review} and ~\ref{ssec:symm_cons} which, while rederiving the empirical tight-binding basic results, also set forth the notation that will be employed throughout the rest of the paper, and help establish a better connection with previous authors' conventions. For our actual double group calculations we use the matrices $D(G)^{\mu}_{is}$ as provided by Shirai~\cite{Shirai1997}, which correspond to the ones tabulated by Onodera and Okazaki~\cite{OnoderaOkazaki1966}.


\section{Results}
\label{sec:application}

In order to illustrate the procedure with a familiar example, and for comparison purposes, we construct in Sec.~\ref{ssec:sg_zb} a 4-band model with no spin included for a zinc blende (point group $T_d$) using an effective bond orbital basis set. Sec.~\ref{ssec:dg} presents the main results of our work, with spin Hamiltonians for various sorts of quantum wells displaying the terms additional to the Rashba and Dresselhaus Hamiltonians. Finally, we show explicitly in Sec.~\ref{ssec:EBOkp} the equivalence between the L\"owdin orbitals (LOs) centered at the primitive cell sites [effective bond orbitals (EBOs)] and the zone center solutions of the \kp\ theory.

The definition of the symbols appearing in the following tables can be found in Equations (\ref{eq:phi_k_melem}) and (\ref{eq:ETB_pars}) in Appendix~\ref{ssec:ETB_review}. The phase of the parameters $E_{ij} ({\bf r})$---the Hamiltonian matrix elements between a state $i$ at the origin and a state $j$ at {\bf r}---has been factored out in the calculations of the tables. Thus, all $E_{ij} ({\bf r})$ are real. Parameters appearing in different tables are of course unrelated, while the procedure in Appendix~\ref{ssec:symm_cons} guarantees a high degree of independence within the parameters in any one table.


\subsection{Single Group}
\label{ssec:sg_zb}

We will study the top valence and lowest conduction bands of a zinc blende (point group $T_d$), which have $\Gamma_5$ and $\Gamma_1$ symmetry at the zone center (we will use the KDWS notation~\cite{KosterDimmockWheeler1963} throughout this paper for the irrep labels). We consider only coupling to the twelve nearest-neighbor sites. Since we are working with effective bond orbitals located at {\em fcc} lattice sites, the nearest neighbors are at ($\oneh$) and its equivalent positions. Table~\ref{tab:Td_single} shows the matrix elements. The use of the arguments in Appendix~\ref{ssec:symm_cons} shows for example that, with our choice of phases, the whole block $E^{\Gamma_5 \Gamma_5}$ [see \refeq{eq:ETB_pars} for meaning] is purely real---which is otherwise trivially obtainable since the Hamiltonian without SO is real and the $p$ orbitals have the same (imaginary) phase---and that $E_{zx} (\oneh) = -E_{xz} (\oneh)$. These results coincide with those of Hass \etal~\cite{HassEhrenreichVelicky1983}, which correct the misprints in Table V of Ref.~\onlinecite{SlaterKoster1954}, or those obtained by the different method of adding a $d$ component to $p$ states in an {\it fcc} lattice~\cite{CartoixaTingMcgill2003c}.

\begin{table}[t]
\centering
\begin{ruledtabular}
\begin{tabular}{cr}
$H^{\Gamma_1 \Gamma_1}_{ss}$({\bf k}) & \parbox[t]{2.9in}{\raggedright $E_{ss} (000) +
   4 E_{ss} (\oneh) \left( \cos \xi \cos \eta + \cos \eta \cos \zeta + \cos \zeta \cos \xi \right)$ } \\
$H^{\Gamma_1 \Gamma_5}_{sx}$({\bf k}) &
  \parbox[t]{2.9in}{\raggedright $4 i E_{sx} (\oneh) \left( \cos \eta + \cos \zeta \right) \sin \xi -
   4 E_{sz} (\oneh) \sin \eta \sin \zeta $ } \\
$H^{\Gamma_5 \Gamma_5}_{xx}$({\bf k}) & \parbox[t]{2.9in}{\raggedright $E_{xx} (000) +
   4 E_{xx} (\oneh) \left( \cos \eta + \cos \zeta \right) \cos \xi +
   4 E_{zz} (\oneh) \cos \eta \cos \zeta $ } \\
$H^{\Gamma_5 \Gamma_5}_{xy}$({\bf k}) & \parbox[t]{2.9in}{\raggedright
  $-4 E_{xy} (\oneh) \sin \xi \sin \eta +
   4 i E_{xz} (\oneh) \left( \cos \xi - \cos \eta \right) \sin \zeta $ }
\end{tabular}
\end{ruledtabular}
\caption{Matrix elements for the zinc blende structure (single group). The definitions $\xi \equiv k_x a/2$, $\eta \equiv k_y a/2$, $\zeta \equiv k_z a/2$, are made. The remaining symbols are defined in Equations (\ref{eq:phi_k_melem}) and (\ref{eq:ETB_pars}). For example, $H^{\Gamma_1 \Gamma_5}_{sx}$({\bf k}) represents the element of the Hamiltonian connecting a state belonging to the $\Gamma_1$ irrep with $s$ symmetry to a state belonging to the $\Gamma_5$ irrep with $x$ symmetry. The parameters $E_{ij} ({\bf r})$ appearing in a table have no relationship with similarly named parameters of a different table.}
\label{tab:Td_single}
\end{table}


\subsection{Double Group}
\label{ssec:dg}

In what follows, we present models of quantum wells (QWs) of different symmetries, including structural inversion asymmetry (SIA)~\cite{BychkovRashba1984:physc} effects only, bulk inversion asymmetry (BIA)~\cite{Dresselhaus1955} only, or both; having in mind the study of in-plane spin transport in two-dimensional electron gases (2DEGs). This has the advantage of yielding less cumbersome expressions than those of the superlattice of the same symmetry, which can be obtained if necessary by setting the matrix elements between supercell instances of different $z$, ${E^{\mu \nu}_{ij} (l,m,n \neq 0)}$, to a finite value. Therefore there will be no $k_z$ terms in the Hamiltonian matrix elements.

Of course, for a numerical zinc blende nanostructure computation the starting point would be a Hamiltonian derived from the bulk matrix elements given at the end of this section. In this context, the following results should rather be taken as an analytical tool for the study of the bands and the spins, and they can be used to test whether calculations from other methods satisfy the symmetry requirements.

\subsubsection{[001] quantum wells}

\paragraph{SIA only}

In this configuration only the Rashba splitting~\cite{BychkovRashba1984:physc} should appear. These quantum wells (QWs) possess $C_{4v}$ symmetry, and an example could be a Si$_{1-x}$Ge$_x$/Si QW with an asymmetric Ge concentration profile. For $C_{4v}$ QWs the CB will transform according to $\Gamma_6$. We construct a model for the CB with on-site and second nearest neighbor coupling, with the assumption that the supercell instances do not couple to each other, and the results are shown in Table~\ref{tab:C4v}.

\begin{figure*}[t]
\centering
\epsfig{file=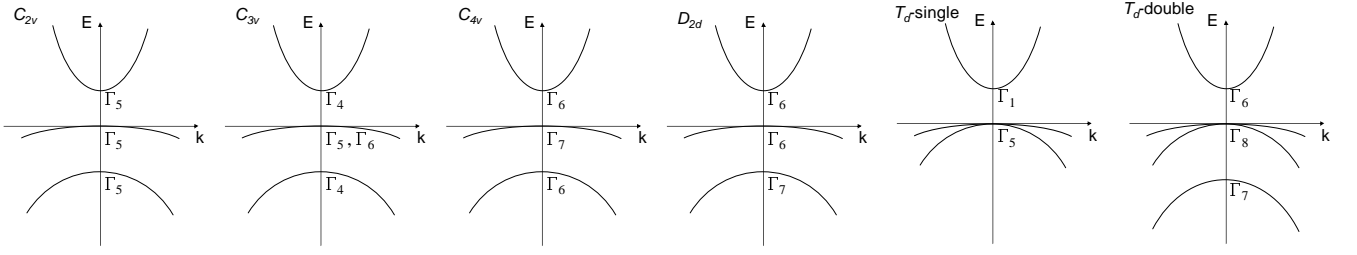, width=0.99\linewidth, clip=}
\caption{Labeling of the irreducible representations (irreps) corresponding to the lowest conduction band and the two highest valence bands, at the zone center, for structures with different point group symmetries. Spin splittings at $k \neq 0$ are omitted for clarity.}
\label{fig:Schemas}
\end{figure*}

\begin{table}[b]
\centering
\begin{ruledtabular}
\begin{tabular}{cr}
$H^{\Gamma_6 \Gamma_6}_{\uparrow \uparrow}$ & \parbox[t]{2.7in}{\raggedright $E_{\uparrow \uparrow} (000) +
   2 E_{\uparrow \uparrow} (100) \left( \cos \xi + \cos \eta \right) +
   4 E_{\uparrow \uparrow} (110) \cos \xi \cos \eta $ } \\
$H^{\Gamma_6 \Gamma_6}_{\uparrow \downarrow}$ & \parbox[t]{2.7in}{\raggedright $
   2 E_{\uparrow \downarrow} (100) \left( i \sin \xi + \sin \eta \right) +
   2 \sqrt{2} E_{\uparrow \downarrow} (110) \left( i \sin \xi \cos \eta + \cos \xi \sin \eta \right) $ }
\end{tabular}
\end{ruledtabular}
\caption{Matrix elements for an [001] structure with SIA only described by the $C_{4v}$ point group. The definitions $\xi \equiv k_x a$ and $\eta \equiv k_y a$ are made.}
\label{tab:C4v}
\end{table}

Expanding again about the $\Gamma$ point and keeping terms of up to third order, we obtain
\begin{multline}
H^{\Gamma_6 \Gamma_6}_{\text{spin}} \approx
   2 a \left[ E_{\uparrow \downarrow} (100) + \sqrt{2} E_{\uparrow \downarrow} (110) \right]
   \left( k_y \sigma_x - k_x \sigma_y \right) + \\
   \left[ E_{\uparrow \downarrow} (100) + \sqrt{2} E_{\uparrow \downarrow} (110) \right] a^3 /3
   \left( - k_y^3 \sigma_x + k_x^3 \sigma_y \right) + \\
   \sqrt{2} a^3 E_{\uparrow \downarrow} (110) k_x k_y \left( - k_x \sigma_x + k_y \sigma_y \right) .
\label{eq:C4v_kp}
\end{multline}

The first term of the right-hand side is, of course, the well-known Rashba Hamiltonian. Here we also derived the two $\mathcal{O}(k^3)$ terms, which are the next order in importance. Up to present there have been calculations that have been carried out to third order in BIA effects but only to linear order in SIA terms~\cite{AverkievGolub1999}. The results from these calculations can be improved and be made consistent within the chosen order of approximation if the higher order SIA terms are included.

\paragraph{BIA only}

\begin{table}[t]
\centering
\begin{ruledtabular}
\begin{tabular}{cr}
$H^{\Gamma_6 \Gamma_6}_{\uparrow \uparrow}$ & \parbox[t]{2.7in}{\raggedright $E_{\uparrow \uparrow} (000) +
   2 E_{\uparrow \uparrow} (100) \left( \cos \xi + \cos \eta \right) +
   4 E_{\uparrow \uparrow} (110) \cos \xi \cos \eta $ } \\
$H^{\Gamma_6 \Gamma_6}_{\uparrow \downarrow}$ & \parbox[t]{2.7in}{\raggedright $
   -2 E_{\uparrow \downarrow} (100) \left( \sin \xi + i \sin \eta \right) +
   2 \sqrt{2} E_{\uparrow \downarrow} (110) \left( \sin \xi \cos \eta + i \cos \xi \sin \eta \right) $ }
\end{tabular}
\end{ruledtabular}
\caption{As Table~\ref{tab:C4v}, but with BIA-only effects ($D_{2d}$).}
\label{tab:D2d}
\end{table}

Here the corresponding point group for a zinc blende [001] QW is $D_{2d}$. The different orientation of the bonds at the interfaces [native inversion asymmetry (NIA)~\cite{VervoortFerreiraVoisin1999}] may lower the symmetry to $C_{2v}$, but this case is equivalent to SIA+BIA, and is treated in the next section. Applying \refeq{eq:dirrep} to the CB ($\Gamma_6$) we obtain the results in Table~\ref{tab:D2d}, and expanding close to $\Gamma$:
\begin{multline}
H^{\Gamma_6 \Gamma_6}_{\text{spin}} \approx
   2 a \left[ E_{\uparrow \downarrow} (100) - \sqrt{2} E_{\uparrow \downarrow} (110) \right]
   \left( - k_x \sigma_x + k_y \sigma_y \right) + \\
   \sqrt{2} a^3 E_{\uparrow \downarrow} (110) k_x k_y \left( - k_y \sigma_x + k_x \sigma_y \right) + \\
   \left[ E_{\uparrow \downarrow} (100) - \sqrt{2} E_{\uparrow \downarrow} (110) \right] a^3 /3
   \left( k_x^3 \sigma_x - k_y^3 \sigma_y \right) .
\label{eq:D2d_kp}
\end{multline}

We recover now the Dresselhaus Hamiltonian for [001] zinc blende quantum wells~\cite{DyakonovKachorovskii1986,EppengaSchuurmans1988:v37,Silva1992} in the first two terms. The middle term can be obtained by the procedure of taking the spin Hamiltonian for bulk zinc blendes [Ref.~\onlinecite{DyakonovPerel1971b} and \refeq{eq:Hamk3} below] and substituting $k_z^2$ for its expectation value in the quantum well~\cite{DyakonovKachorovskii1986}. However, the last term cannot be obtained in this fashion,~\cite{endnote45} which shows the power of our proposed approach. The symbols $E_{\uparrow \uparrow} (100)$, $E_{\uparrow \downarrow} (110)$, etc., have no relationship with their analogous in the other cases, and the reference to the corresponding point group has been dropped to lighten the notation.

\paragraph{SIA+BIA}

\begin{table}[t]
\centering
\begin{ruledtabular}
\begin{tabular}{cr}
$H^{\Gamma_5 \Gamma_5}_{\uparrow \uparrow}$ & \parbox[t]{2.7in}{\raggedright $E_{\uparrow \uparrow} (000) +
   2 E_{\uparrow \uparrow} (100) \cos \xi + 2 E_{\uparrow \uparrow} (010) \cos \eta +
   4 E_{\uparrow \uparrow} (110) \cos \xi \cos \eta $ } \\
$H^{\Gamma_5 \Gamma_5}_{\uparrow \downarrow}$ & \parbox[t]{2.7in}{\raggedright $
   2 E_{\uparrow \downarrow} (100) i \sin \xi - 2 E_{\uparrow \downarrow} (010) \sin \eta -
   4 \Im \left[ E_{\uparrow \downarrow} (110) \right] \cos \xi \sin \eta +
   4 \Re \left[ E_{\uparrow \downarrow} (110) \right] i \sin \xi \cos \eta $ }
\end{tabular}
\end{ruledtabular}
\caption{As Table~\ref{tab:C4v}, but with SIA and BIA effects ($C_{2v}$). The phase of $E_{\uparrow \downarrow} (110)$ has not been factored out as it is not determined by symmetry.}
\label{tab:C2v}
\end{table}

The corresponding point group for these structures is $C_{2v}$, and the conduction band is associated with the $\Gamma_5$ irrep. Table~\ref{tab:C2v} shows the results, and the spin Hamiltonian is
\begin{multline}
H^{\Gamma_5 \Gamma_5}_{\text{spin}} \approx
   -2 a \left\{
   \left[ E_{\uparrow \downarrow} (010) + 2 \Im \left[ E_{\uparrow \downarrow} (110) \right] \right] k_y \sigma_x +
   \right. \\
   \left.
   \left[ E_{\uparrow \downarrow} (100) + 2 \Re \left[ E_{\uparrow \downarrow} (110) \right] \right] k_x \sigma_y
   \right\} + \\
   a^3 /3 \left\{
   \left[ E_{\uparrow \downarrow} (010) + 2 \Im \left[ E_{\uparrow \downarrow} (110) \right] \right] k_y^3 \sigma_x +
   \right. \\
   \left[ E_{\uparrow \downarrow} (100) + 2 \Re \left[ E_{\uparrow \downarrow} (110) \right] \right]
   k_x^3 \sigma_y + \\
   \left.
   6 \Im \left[ E_{\uparrow \downarrow} (110) \right] k_x^2 k_y \sigma_x +
   6 \Re \left[ E_{\uparrow \downarrow} (110) \right] k_y^2 k_x \sigma_y   \right\} .
\end{multline}

This Hamiltonian can be transformed into a more familiar form by making the substitutions $E_{\uparrow \downarrow} (010) \rightarrow E_{\text{BIA}} + E_{\text{SIA}}$, $E_{\uparrow \downarrow} (100) \rightarrow E_{\text{BIA}} - E_{\text{SIA}}$ and $E_{\uparrow \downarrow} (110) \rightarrow \left(E_{\text{BIA,110}} + E_{\text{SIA,110}} \right)/2 + i \left(E_{\text{BIA,110}} - E_{\text{SIA,110}} \right)/2$ and reverting to the original axes definitions for the zinc blende structure:
\begin{multline}
H^{\Gamma_5 \Gamma_5}_{\text{spin}} \approx
   2 a \left[ E_{\text{SIA}} - 2 E_{\text{SIA,110}} \right] \left( - k_y \sigma_x + k_x \sigma_y \right) + \\
   2 a \left[ E_{\text{BIA}} + 2 E_{\text{BIA,110}} \right] \left( - k_x \sigma_x + k_y \sigma_y \right) + \\
   a^3 /6 \left\{
   \left[ E_{\text{SIA}} - 4 E_{\text{SIA,110}} \right] \left( k_y^3 \sigma_x - k_x^3 \sigma_y \right) + \right. \\
   \left[ E_{\text{BIA}} + 4 E_{\text{BIA,110}} \right] \left( k_x^3 \sigma_x - k_y^3 \sigma_y \right) + \\
   3 E_{\text{SIA}} k_x k_y \left( k_x \sigma_x - k_y \sigma_y \right) +
   \left. 3 E_{\text{BIA}} k_x k_y \left( k_y \sigma_x - k_x \sigma_y \right)   \right\} ,
\end{multline}
which contains the same functional dependence as the sum of the separated SIA- and BIA-only cases, Eqs.~(\ref{eq:C4v_kp}) and (\ref{eq:D2d_kp}).

\subsubsection{[110] quantum wells}
\label{sssec:110_QWs}

\paragraph{SIA only}

The symmetry group is $C_{2v}$, so the results are essentially the same as for [001] heterostructures with BIA+SIA. It is worth noting that, because of the reduced symmetry with respect to SIA-[001], there appear additional Dresselhaus-like terms although for SIA-only one would expect only Rashba terms.

\begin{figure}[t]
\centering
\epsfig{file=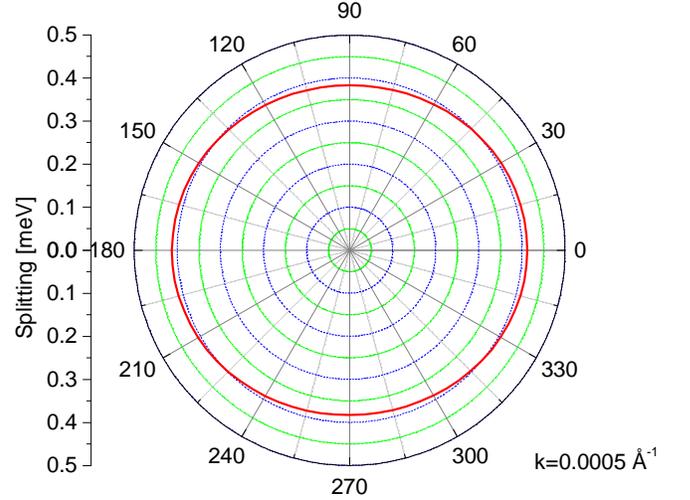, width=0.99\linewidth, clip=}
\caption{Polar plot of the spin splitting in the CB for a [110] 16/16 AlSb/GaSb/InAs/AlSb QW with SIA only calculated with \kp. The anisotropy of the splitting is due to the interplay of Rashba- and Dresselhaus-like terms in the spin Hamiltonian, although the simulation only accounts for SIA effects.}
\label{fig:NoBIA_110_polar}
\end{figure}

We have verified these predictions with the help of \kp\ code where BIA effects can be turned on and off~\cite{CartoixaTingMcgill2003:nanotech} for a [110] 16/16 AlSb/GaSb/InAs/AlSb QW (see Fig.~\ref{fig:NoBIA_110_polar}). We see for this case that the splitting presents a small degree of anisotropy due to the interplay of Rashba- and Dresselhaus-like terms in the spin Hamiltonian, although the simulation only accounts for SIA effects. The axes of constructive and destructive Rashba--Dresselhaus interference are rotated by $\pi/4$ with respect to the [001] BIA+SIA case~\cite{Silva1992} because here the planes of symmetry coincide with the usual choice of $x$ and $y$ axes, while in the [001] the reflection planes bisect the $x$ and $y$ axes.

\paragraph{BIA only}

The symmetry group is again $C_{2v}$, but now the twofold axis is in-plane, whereas in the previous case it was along the growth direction. This changes the symmetry considerations considerably. The results for a second-nearest-neighbor model are displayed in Table~\ref{tab:C2v_110}.

\begin{table}[t]
\centering
\begin{ruledtabular}
\begin{tabular}{cr}
$H^{\Gamma_5 \Gamma_5}_{\uparrow \uparrow}$ & \parbox[t]{2.7in}{\raggedright $ E_{\uparrow \uparrow} (000) +
   2 \Re \left[ E_{\uparrow \uparrow} (100) \right] \cos \xi -
   2 \Im \left[ E_{\uparrow \uparrow} (100) \right] \sin \xi +
   2 E_{\uparrow \uparrow} (010) \cos \eta +
   4 \Re \left[ E_{\uparrow \uparrow} (110) \right] \cos \xi \cos \eta -
   4 \Im \left[ E_{\uparrow \uparrow} (110) \right] \sin \xi \cos \eta
   $ } \\
$H^{\Gamma_5 \Gamma_5}_{\uparrow \downarrow}$ & \parbox[t]{2.7in}{\raggedright $
   0 $ }
\end{tabular}
\end{ruledtabular}
\caption{Matrix elements for an [110] structure with BIA only described by the $C_{2v}$ point group. The definitions $\xi \equiv k_x a_x$ and $\eta \equiv k_y a_y$ are made.}
\label{tab:C2v_110}
\end{table}

An interesting feature that is recovered is that, for $k_z =0$ (or for a quantum well), the spin will align or antialign along the growth direction. This result still holds even when all the neighbors are included: a general ${\bf k}$ point in the well Brillouin zone has $C_s$ symmetry (reflection with respect to the well plane), which confines the spin to lie perpendicular to the plane. This makes symmetric [110] structures have unusually long spin lifetimes~\cite{DyakonovKachorovskii1986,OhnoTerauchiAdachi1999} for spins along the growth direction because of the suppression of the D'yakonov-Perel' (DP) spin relaxation mechanism~\cite{DyakonovPerel1971b}. Then, the spin Hamiltonian takes the simple form
\begin{multline}
H^{\Gamma_5 \Gamma_5}_{\text{spin}} \approx
   -2 a_x \Im \left[ E_{\uparrow \uparrow} (100) + 2 E_{\uparrow \uparrow} (110) \right] k_x \sigma_z + \\
   a_x/3 \left\{
   a_x^2 \Im \left[ E_{\uparrow \uparrow} (100) + 2 E_{\uparrow \uparrow} (110) \right] k_x^2 + \right. \\
   \left. a_y^2 6 \Im \left[ E_{\uparrow \uparrow} (110) \right] k_y^2  \right\} k_x \sigma_z .
\end{multline}

\paragraph{SIA+BIA}

In this case the symmetry will be lowered to $C_s$ (a single reflection plane), and the above arguments will not be able to put any restrictions on the spins. This is further understood if we combine the BIA-only results, which keep the spins pointing along the growth direction, with the SIA-only results, which tend to keep the spins perpendicular to the growth axis. Thus, the spins will point in a general direction dependent on the particular structure under study.

\subsubsection{[111] quantum wells}

For BIA-only, these quantum wells transform according to the $C_{3v}$ point group~\cite{KitaevPanfilovTronc1997}; consistent of the identity, two threefold rotations about the growth axis and three reflection planes separated by 120\deg\ that contain the threefold axis. It is easy to see by inspection that the inclusion of SIA does not reduce the symmetry with respect to the BIA-only case. Thus, here it is only necessary to derive the TB Hamiltonian for $C_{3v}$ structures, which is shown in Table~\ref{tab:C3v}.

\begin{table}[t]
\centering
\begin{ruledtabular}
\begin{tabular}{cr}
$H^{\Gamma_4 \Gamma_4}_{\uparrow \uparrow}$ & \parbox[t]{2.7in}{\raggedright $ E_{\uparrow \uparrow} (000) +
   2 \Re \left[ E_{\uparrow \uparrow} (010) \right] \left( 2 \cos \sqrt{3} \xi/2 \cos \eta/2 + \cos \eta \right) +
   2 \Im \left[ E_{\uparrow \uparrow} (010) \right] \left( 2 \cos \sqrt{3} \xi/2 \sin \eta/2 - \sin \eta \right) +
   2 E_{\uparrow \uparrow} (\sqrt{3}00)
   \left[ \cos \sqrt{3} \xi + \cos (\sqrt{3} \xi/2 - 3\eta/2) \right. + \left. \cos (\sqrt{3} \xi/2 + 3\eta/2) \right] +
   $ } \\
$H^{\Gamma_4 \Gamma_4}_{\uparrow \downarrow}$ & \parbox[t]{2.7in}{\raggedright $
   - 2 E_{\uparrow \downarrow} (010) ( \cos \sqrt{3} \xi/2 \sin \eta/2 + \sin \eta +
   i \sqrt{3} \sin \sqrt{3} \xi/2 \cos \eta/2 ) +
   i E_{\uparrow \downarrow} (\sqrt{3}00)
   \left[ 2 \sin \sqrt{3} \xi + \right.
   (1 + i \sqrt{3}) \sin \left( \sqrt{3} \xi/2 - 3\eta/2 \right) +
   \left. (1 - i \sqrt{3}) \sin (\sqrt{3} \xi/2 + 3\eta/2) \right]      $ }
\end{tabular}
\end{ruledtabular}
\caption{Matrix elements for an [111] structure described by the $C_{3v}$ point group. The definitions $\xi \equiv k_x a$ and $\eta \equiv k_y a$ are made, and the primitive vectors are taken to be $a(0,1)$ and $a(\sqrt{3}/2,1/2)$.}
\label{tab:C3v}
\end{table}

The expansion about the $\Gamma$ point yields a spin Hamiltonian:
\begin{multline}
H^{\Gamma_4 \Gamma_4}_{\text{spin}} \approx
   3 a \left[ \sqrt{3} E_{\uparrow \downarrow} (\sqrt{3}00) - E_{\uparrow \downarrow} (010) \right]
   \left( k_y \sigma_x - k_x \sigma_y \right) + \\
   a^3/8 \left\{
   \left[ 3 E_{\uparrow \downarrow} (010) - 9 \sqrt{3} E_{\uparrow \downarrow} (\sqrt{3}00) \right]
   k^2 \left( k_y \sigma_x - k_x \sigma_y \right) + \right.\\
   \left. 2 \Im \left[ E_{\uparrow \uparrow} (010) \right] ( -3 k_x^2 + k_y^2 ) k_y \sigma_z  \right\} .
\end{multline}
Thus, no matter for what situation in [111] QWs, we only have---at $\mathcal{O}(k)$---Rashba-like terms, as shown previously with a different method~\cite{CartoixaTingChang2005}. This is important because the Rashba contribution can be tuned through the action of a gate bias, and it is in principle possible to make the first order spin splitting vanish, suppressing the DP spin relaxation mechanism for all components at the same time~\cite{CartoixaTingChang2005}.

\subsubsection{Zinc blendes with effective bond orbitals}

Here we will construct the model equivalent to the $H^{\Gamma_5 \Gamma_5}$ block in Sec.~\ref{ssec:sg_zb}, but with spin. Thus, it will include a heavy- and light-hole $\Gamma_8$ and a split-off $\Gamma_7$ set of states. We provide the symmetry-constrained matrices $E^{\Gamma_i \Gamma_j}$, as they are the primary quantities in nanostructures:
\begin{equation}
E^{\Gamma_8 \Gamma_8} (\oneh) =
\begin{pmatrix}
 E_{33} & \omega^{\ast} E_{31} & i E_{3\bar{1}} & \omega E_{3\bar{3}} \\
-\omega E_{31} & E_{11} & \omega^{\ast} E_{1\bar{1}} & i E_{3\bar{1}} \\
-i E_{3\bar{1}} & -\omega E_{1\bar{1}} & E_{11} & \omega^{\ast} E_{31} \\
-\omega^{\ast} E_{3\bar{3}} & -i E_{3\bar{1}} & -\omega E_{31} & E_{33}
\end{pmatrix} ,
\label{eq:Emelem_init}
\end{equation}
\begin{equation}
E^{\Gamma_8 \Gamma_7} (\oneh) =
\begin{pmatrix}
\omega^{\ast} E_{3 \uparrow} & i E_{\bar{3} \uparrow} \\
E_{1 \uparrow} & \omega^{\ast} E_{\bar{1} \uparrow} \\
\omega E_{\bar{1} \uparrow} & -E_{1 \uparrow} \\
i E_{\bar{3} \uparrow} & \omega E_{3 \uparrow}
\end{pmatrix} ,
\end{equation}
\begin{equation}
E^{\Gamma_7 \Gamma_8} (\oneh) =
\begin{pmatrix}
-\omega E_{3 \uparrow} & E_{1 \uparrow} & -\omega^{\ast} E_{\bar{1} \uparrow} & -i E_{\bar{3} \uparrow} \\
-i E_{\bar{3} \uparrow} & -\omega E_{\bar{1} \uparrow} & E_{1 \uparrow} & -\omega^{\ast} E_{3 \uparrow}
\end{pmatrix} ,
\end{equation}
and
\begin{equation}
E^{\Gamma_7 \Gamma_7} (\oneh) =
\begin{pmatrix}
E_{\uparrow \uparrow} & \omega^{\ast} E_{\uparrow \downarrow} \\
-\omega E_{\uparrow \downarrow} & E_{\uparrow \uparrow}
\end{pmatrix} ,
\label{eq:Emelem_end}
\end{equation}
where $\omega \equiv e^{i \pi /4}$, $\bar{l} \equiv -l$, $l$ refers to $\ket{\phi^{\Gamma_8}_{l/2}}$ and $\uparrow$ ($\downarrow$) refers to $\ket{\phi^{\Gamma_7}_{1/2}}$ ($\ket{\phi^{\Gamma_7}_{-1/2}}$), and the phase factors are explicitly shown so that all parameters $E_{ij}$ are real. Thus, using symmetry operations, the 36 initial matrix elements are reduced to 12 independent parameters. The matrix elements $E^{\Gamma_i \Gamma_j} (G{\bf r}_j)$ for the remaining nearest-neighbor sites are obtained combining Eqs.~(\ref{eq:Emelem_init})-(\ref{eq:Emelem_end}) and (\ref{eq:Eirrep}).

Now we can construct the bulk tight-binding matrix elements, which will be the foundation for further analysis of the meaning of the parameters. Table~\ref{tab:Td_double} shows the computed matrix elements from Eqs.~(\ref{eq:dirrep}) and (\ref{eq:sym_Hmelem}). The remaining matrix elements can be obtained from the hermiticity of the Hamiltonian in the Bloch sum representation. We can obtain information about the physical effects that are included in this model by expanding the matrix elements~\cite{Chang1988} about ${\bf k}=0$ and comparing to \kp\ results~\cite{Kane1966}.

\begin{table*}[t]
\centering
\renewcommand{\arraystretch}{1.3}
\begin{ruledtabular}
\begin{tabular}{cr}
$H^{\Gamma_8 \Gamma_8}_{33}$({\bf k}) & \parbox[t]{6in}{\raggedright $E_{33} (000) +
   [ 3E_{11} (\oneh) + E_{33} (\oneh) ] \left( \cos \xi \cos \eta + \cos \eta \cos \zeta + \cos \zeta \cos \xi \right) +
   3 E_{33} (\oneh) \cos \xi \cos \eta + \newline
   \frac{1}{\sqrt{2}} [3E_{1\bar{1}} (\oneh) + 2\sqrt{3} E_{31} (\oneh) - E_{3\bar{3}} (\oneh)]
   \left( \cos \xi - \cos \eta \right) \sin \zeta $ } \\
$H^{\Gamma_8 \Gamma_8}_{11}$({\bf k}) & \parbox[t]{6in}{\raggedright $E_{33} (000) +
   [ 3E_{33} (\oneh) + E_{11} (\oneh) ] \left( \cos \xi \cos \eta + \cos \eta \cos \zeta + \cos \zeta \cos \xi \right) +
   3 E_{11} (\oneh) \cos \xi \cos \eta - \newline
   \frac{1}{\sqrt{2}} [E_{1\bar{1}} (\oneh) - 2\sqrt{3} E_{31} (\oneh) - 3 E_{3\bar{3}} (\oneh)]
   \left( \cos \xi - \cos \eta \right) \sin \zeta$ } \\
$H^{\Gamma_8 \Gamma_8}_{31}$({\bf k}) & \parbox[t]{6in}{\raggedright $ - \left[ \sqrt{2} E_{31} (\oneh) +
   \sqrt{3/2} ( E_{1\bar{1}} (\oneh) + E_{3\bar{3}} (\oneh) ) \right] \left( \sin \xi + i \sin \eta \right) \cos \zeta +
   2 \sqrt{2} E_{31} (\oneh) \left( \sin \xi \cos \eta + i \cos \xi \sin \eta \right) +
   4 E_{3\bar{1}} (\oneh) \left( \sin \xi - i \sin \eta \right) \sin \zeta
   $ } \\
$H^{\Gamma_8 \Gamma_8}_{3\bar{1}}$({\bf k}) & \parbox[t]{6in}{\raggedright $ \left[ - \sqrt{2} E_{31} (\oneh) +
   \sqrt{3/2} ( E_{1\bar{1}} (\oneh) + E_{3\bar{3}} (\oneh) ) \right] \left( \cos \xi + \cos \eta \right) \sin \zeta +
   \sqrt{3} ( E_{11} (\oneh) - E_{33} (\oneh) ) \left( \cos \xi - \cos \eta \right) \cos \zeta -
   4 i E_{3\bar{1}} (\oneh) \sin \xi \sin \eta
   $ } \\
$H^{\Gamma_8 \Gamma_8}_{3\bar{3}}$({\bf k}) & \parbox[t]{6in}{\raggedright $ \left[ -3 E_{1\bar{1}} (\oneh) / \sqrt{2} +
   \sqrt{6} E_{31} (\oneh) + E_{3\bar{3}} (\oneh)  / \sqrt{2} \right] \left( \sin \xi - i \sin \eta \right) \cos \zeta +
   2 \sqrt{2} E_{3\bar{3}} (\oneh) \left( - \sin \xi \cos \eta + i \cos \xi \sin \eta \right)
   $ } \\
$H^{\Gamma_8 \Gamma_8}_{1\bar{1}}$({\bf k}) & \parbox[t]{6in}{\raggedright $ - \left[ E_{1\bar{1}} (\oneh) / \sqrt{2} +
   \sqrt{6} E_{31} (\oneh) - 3E_{3\bar{3}} (\oneh)  / \sqrt{2} \right] \left( \sin \xi + i \sin \eta \right) \cos \zeta +
   2 \sqrt{2} E_{1\bar{1}} (\oneh) \left( \sin \xi \cos \eta + i \cos \xi \sin \eta \right)
   $ } \\
\multicolumn{2}{l}{
    $H^{\Gamma_8 \Gamma_8}_{\bar{3}\bar{3}} (k_x,k_y,k_z) = H^{\Gamma_8 \Gamma_8}_{33} (k_x,-k_y,-k_z)$;
    $H^{\Gamma_8 \Gamma_8}_{\bar{1}\bar{1}} (k_x,k_y,k_z) = H^{\Gamma_8 \Gamma_8}_{11} (k_x,-k_y,-k_z)$
} \\
\multicolumn{2}{l}{
    $H^{\Gamma_8 \Gamma_8}_{1\bar{3}} (k_x,k_y,k_z) = {H^{\Gamma_8 \Gamma_8}_{3\bar{1}}}^\ast (k_x,-k_y,-k_z)$;
    $H^{\Gamma_8 \Gamma_8}_{\bar{1}\bar{3}} (k_x,k_y,k_z) = {H^{\Gamma_8 \Gamma_8}_{31}}^\ast (k_x,-k_y,-k_z)$
} \\
$H^{\Gamma_7 \Gamma_7}_{\uparrow \uparrow}$({\bf k}) & \parbox[t]{6in}{\raggedright $E_{\uparrow \uparrow} (000) +
   4 E_{\uparrow \uparrow} (\oneh) \left( \cos \xi \cos \eta + \cos \eta \cos \zeta + \cos \zeta \cos \xi \right) +
   2 \sqrt{2} E_{\uparrow \downarrow} (\oneh) \left( \cos \xi - \cos \eta \right) \sin \zeta
   $ } \\
$H^{\Gamma_7 \Gamma_7}_{\uparrow \downarrow}$({\bf k}) & \parbox[t]{6in}{\raggedright $
   2 \sqrt{2} E_{\uparrow \downarrow} (\oneh) \left[ \left( \cos \eta - \cos \zeta \right) \sin \xi -
   i \left( \cos \zeta - \cos \xi \right) \sin \eta \right]
   $ } \\
\multicolumn{2}{l}{
    $H^{\Gamma_7 \Gamma_7}_{\downarrow \downarrow} (k_x,k_y,k_z) =
    H^{\Gamma_7 \Gamma_7}_{\uparrow \uparrow} (k_x,-k_y,-k_z)$
} \\
$H^{\Gamma_8 \Gamma_7}_{3 \uparrow}$({\bf k}) & \parbox[t]{6in}{\raggedright $
   2 \sqrt{2} E_{3 \uparrow} (\oneh) \left( \sin \xi \cos \eta + i \cos \xi \sin \eta \right) -
   ( \sqrt{2} E_{3 \uparrow} (\oneh) + \sqrt{6} E_{\bar{1} \uparrow} (\oneh) ) \left( \sin \xi + i \sin \eta \right)
   \cos \zeta +
   2 E_{\bar{3} \uparrow} (\oneh) \left( \sin \xi - i \sin \eta \right) \sin \zeta
   $ } \\
$H^{\Gamma_8 \Gamma_7}_{1 \uparrow}$({\bf k}) & \parbox[t]{6in}{\raggedright $
   4 E_{1 \uparrow} (\oneh) \cos \xi \cos \eta -
   2 E_{1 \uparrow} (\oneh) \left( \cos \xi + \cos \eta \right) \cos \zeta -
   ( \sqrt{6} E_{3 \uparrow} (\oneh) + \sqrt{2} E_{\bar{1} \uparrow} (\oneh) ) \left( \cos \xi - \cos \eta \right)
   \sin \zeta
   $ } \\
$H^{\Gamma_8 \Gamma_7}_{\bar{1} \uparrow}$({\bf k}) & \parbox[t]{6in}{\raggedright $
   2 \sqrt{2} E_{\bar{1} \uparrow} (\oneh) \left( - \sin \xi \cos \eta + i \cos \xi \sin \eta \right) +
   ( \sqrt{6} E_{3 \uparrow} (\oneh) - \sqrt{2} E_{\bar{1} \uparrow} (\oneh) ) \left( \sin \xi - i \sin \eta \right)
   \cos \zeta +
   2 \sqrt{3} E_{\bar{3} \uparrow} (\oneh) \left( \sin \xi + i \sin \eta \right) \sin \zeta
   $ } \\
$H^{\Gamma_8 \Gamma_7}_{\bar{3} \uparrow}$({\bf k}) & \parbox[t]{6in}{\raggedright $
   -4 i E_{\bar{3} \uparrow} (\oneh) \sin \xi \sin \eta -
   2 \sqrt{3} E_{1 \uparrow} (\oneh) \left( \cos \xi - \cos \eta \right) \cos \zeta +
   ( \sqrt{2} E_{3 \uparrow} (\oneh) - \sqrt{6} E_{\bar{1} \uparrow} (\oneh) ) \left( \cos \xi + \cos \eta \right)
   \sin \zeta
   $ } \\
\multicolumn{2}{l}{
    $H^{\Gamma_8 \Gamma_7}_{\bar{3}\downarrow} (k_x,k_y,k_z) = -H^{\Gamma_8 \Gamma_7}_{3\uparrow} (k_x,-k_y,-k_z)$;
    $H^{\Gamma_8 \Gamma_7}_{\bar{1}\downarrow} (k_x,k_y,k_z) = -H^{\Gamma_8 \Gamma_7}_{1\uparrow} (k_x,-k_y,-k_z)$;
} \\
\multicolumn{2}{l}{
    $H^{\Gamma_8 \Gamma_7}_{1\downarrow} (k_x,k_y,k_z) = -H^{\Gamma_8 \Gamma_7}_{\bar{1}\uparrow} (k_x,-k_y,-k_z)$;
    $H^{\Gamma_8 \Gamma_7}_{3\downarrow} (k_x,k_y,k_z) = -H^{\Gamma_8 \Gamma_7}_{\bar{3}\uparrow} (k_x,-k_y,-k_z)$;
} \\
\end{tabular}
\end{ruledtabular}
\caption{Matrix elements for the zinc blende structure (double group). The remaining matrix elements can be obtained from the hermiticity of the Hamiltonian in the Bloch sum representation. The definitions $\xi \equiv k_x a/2$, $\eta \equiv k_y a/2$, $\zeta \equiv k_z a/2$, are made.}
\label{tab:Td_double}
\end{table*}

We focus first on the heavy hole (HH) and light hole (LH) bands, given by the $\Gamma_7$ irrep. ETB implementations based on atomic or bond orbitals with the addition of on-site spin-orbit~\cite{Chadi1977,CartoixaTingMcgill2003c} yield an incorrect intraband splitting proportional to $k^3$ along [110], while Dresselhaus showed~\cite{Dresselhaus1955} from symmetry arguments that it should be linear with $k$. Our current method solves this problem. From \kp\ theory we know that a parameter $C$ describes the linear splitting~\cite{Kane1966}. Looking, for example, at the $H^{\Gamma_8 \Gamma_8}_{\bar{3}3}$ matrix element and comparing to the corresponding \kp\ matrix element, $-\sqrt{3}C (k_x +i k_y)/2$, we can identify
\begin{equation}
C = \sqrt{2} a \left[ \sqrt{3} E_{1\bar{1}} (\oneh) - 2 E_{31} (\oneh) + \sqrt{3} E_{3\bar{3}} (\oneh) \right] .
\end{equation}
If the parameters in $C$ were calculated from single group theory, we would obtain $E_{1\bar{1}} (\oneh) = - 2\sqrt{2} E_{xz} (\oneh) /3$, $E_{31} (\oneh) = - \sqrt{2/3} E_{xz} (\oneh)$ and $E_{3\bar{3}} (\oneh) = 0$, yielding $C=0$ and, therefore, no linear splitting in the HH--LH bands.

\begin{figure}[t]
\centering
\epsfig{file=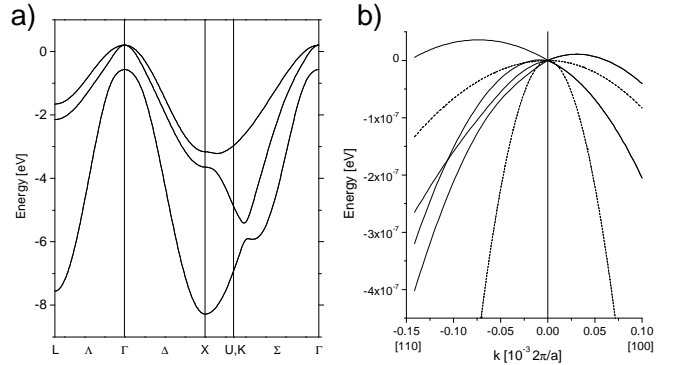, width=0.99\linewidth, clip=}
\caption{a) Full zone HH, LH and SO bands for GaSb. b) Bands very close to the zone center along [100] and [110] directions calculated with the present method (solid lines), featuring the linear splitting, and calculated with the method of Ref.~\onlinecite{CartoixaTingMcgill2003c} (dashed lines).}
\label{fig:GaSb}
\end{figure}

Figure~\ref{fig:GaSb} shows the calculated HH, LH and SO bands for bulk GaSb. For actual calculations to show the inclusion of the effect, we set the double group parameters to the single group values as determined from the Effective Bond Orbital Model (EBOM) method~\cite{Chang1988}, with the addition of $E_{3\bar{3}} (\oneh) = \sqrt{2/3} C /a$. We have taken $C=0.7$ meV$\cdot$\AA\ as calculated by Cardona \etal\cite{CardonaChristensenFasol1988} Plot (a) shows the full zone bands. The inclusion of $E_{3\bar{3}} (\oneh)$ makes no appreciable change to the bands at the scale of the plot (\ie\ no linear splitting is visible at this scale). Plot (b) is a closer look at the HH and LH bands at the top of the valence band comparing the method presented here (solid lines) and a previous EBOM implementation allowing for bulk inversion asymmetry~\cite{CartoixaTingMcgill2003c} (dashed lines). Only the present method reproduces the linear splittings, in accordance with the predictions from group theory~\cite{Dresselhaus1955}, showing that the linear splitting arises from off-site SO interactions, which were not included in Ref.~\onlinecite{CartoixaTingMcgill2003c}. In this particular case, the scale of the energy is tiny, and it makes little sense to make any claim when such small energy scales are involved. Nevertheless, this example shows that it is indeed possible to describe correctly {\em all} spin-orbit effects within a tight-binding framework and, moreover, the effects of a finite $C$ have been observed experimentally for GaSb in hole transport measurements~\cite{MathurJain1979,HellerHamerly1985}. We show below more examples (see, for example, Sec.~\ref{sssec:110_QWs}~{\it a}) with potentially bigger magnitudes.

We now turn our attention to the lowest conduction band in zinc blendes, which is associated to the $\Gamma_6$ irrep. All the symmetry-equivalent points of $(\oneh)$ are spanned by the proper rotations of $T_d$, which under $\Gamma_6$ correspond to the same matrices as $\Gamma_7$. Therefore, the $H^{\Gamma_7 \Gamma_7}$ matrix elements remain unchanged under the substitution $\Gamma_7 \rightarrow \Gamma_6$, leading to a full zone spin Hamiltonian
\begin{equation}
H^{\Gamma_6 \Gamma_6}_{\text{spin}} =
   2 \sqrt{2} E^{\Gamma_6 \Gamma_6}_{\uparrow \downarrow} (\oneh)
   \left[ \sigma_z \left( \cos \xi - \cos \eta \right) \sin \zeta + c.p. \right] ,
\end{equation}
where $c.p.$ stands for cyclic permutations and which, after expanding to lowest order about the zone center, allows us to recover the $k^3$ Hamiltonian~\cite{DyakonovPerel1971b}
\begin{equation}
H^{\Gamma_6 \Gamma_6}_{\text{spin}} \approx
   \gamma_c
   \left[ \left( k_x^2 - k_y^2 \right) k_z \sigma_z + c.p. \right] ,
\label{eq:Hamk3}
\end{equation}
with $\gamma_c \equiv a^3 / (2 \sqrt{2}) E^{\Gamma_6 \Gamma_6}_{\uparrow \downarrow} (\oneh)$ and $\sigma_i$ the Pauli spin matrices.

\subsubsection{Zinc blendes with quasi-atomic orbitals}

Within ETB, a more common approach for the study of zinc blendes is the use of orbitals assigned to atomic positions~\cite{SlaterKoster1954,ChadiCohen1975,Harrison,VoglHjalmarsonDow1983} rather than Bravais lattice sites. This increases the spatial resolution and the size of the basis set. The anion is assumed to have ${\bf d}_a=0$, while the cation is at ${\bf d}_c=a(1,1,1)/4$. The blocks for the first-neighbor overlaps for VB states are then given by
\begin{equation}
E^{\Gamma^a_7 \Gamma^c_7} (\oneq) =
\begin{pmatrix}
E_{\uparrow \uparrow} & 0 \\
0 & E_{\uparrow \uparrow}
\end{pmatrix} ,
\end{equation}
\begin{multline}
E^{\Gamma^a_7 \Gamma^c_8} (\oneq) = \\
\begin{pmatrix}
\omega E_{\uparrow 3} & 0 & -\omega^{\ast} \sqrt{3} E_{\uparrow 3} & i \sqrt{2} E_{\uparrow 3} \\
i \sqrt{2} E_{\uparrow 3} & -\omega \sqrt{3} E_{\uparrow 3} & 0 & \omega^{\ast} E_{\uparrow 3}
\end{pmatrix} ,
\end{multline}
\begin{equation}
E^{\Gamma^a_8 \Gamma^c_7} (\oneq) =
\begin{pmatrix}
\omega^{\ast} E_{3 \uparrow} & -i \sqrt{2} E_{3 \uparrow} \\
0 & -\omega^{\ast} \sqrt{3} E_{3 \uparrow} \\
-\omega \sqrt{3} E_{3 \uparrow} & 0 \\
-i \sqrt{2} E_{3 \uparrow} & \omega E_{3 \uparrow}
\end{pmatrix} ,
\end{equation}
and
\begin{widetext}
\begin{multline}
E^{\Gamma^a_8 \Gamma^c_8} (\oneq) = \\
\begin{pmatrix}
 E_{33} & -\omega^{\ast} [ \sqrt{2} E_{3\bar{1}} + \sqrt{3} E_{3\bar{3}} ] & i E_{3\bar{1}} & \omega E_{3\bar{3}} \\
-\omega [ \sqrt{2} E_{3\bar{1}} + E_{3\bar{3}} / \sqrt{3} ] & E_{33} & \omega^{\ast} E_{3\bar{3}} & i E_{3\bar{1}} + 2 i \sqrt{2/3} E_{3\bar{3}} \\
-i E_{3\bar{1}} - 2 i \sqrt{2/3} E_{3\bar{3}} & -\omega E_{3\bar{3}} & E_{33} & \omega^{\ast} [ \sqrt{2} E_{3\bar{1}} + E_{3\bar{3}} / \sqrt{3} ] \\
-\omega^{\ast} E_{3\bar{3}} & -i E_{3\bar{1}} & \omega [ \sqrt{2} E_{3\bar{1}} + \sqrt{3} E_{3\bar{3}} ] & E_{33}
\end{pmatrix} ,
\label{eq:EmelemQAO}
\end{multline}
\end{widetext}
and $E^{\Gamma^c_i \Gamma^a_j} (\moneq)$ can be obtained from
\begin{equation}
{}^{\alpha' \alpha} \! {E^{\mu'\mu}_{i'i}} (-{\bf R}_j - {\bf d}_{\alpha'} + {\bf d}_\alpha) =
{}^{\alpha \alpha'} \! {E^{\mu\mu'}_{ii'}}^{\ast} ( {\bf R}_j + {\bf d}_{\alpha'} - {\bf d}_\alpha) ,
\label{eq:Trans_sym_ac}
\end{equation}
with $\alpha,\alpha'=a,c$ (anion or cation). The spinless model is recovered setting $E_{33} = E_{\uparrow \uparrow} = E_{xx}$, $E_{3\bar{1}} = E_{3 \uparrow} = E_{\uparrow 3} = E_{xy}/\sqrt{3}$ and $E_{3\bar{3}} = 0$. The non-trivial Hamiltonian matrix elements in the bulk orbital representation are given in Table~\ref{tab:Td_double_QAO}.

\begin{table}[t]
\centering
\renewcommand{\arraystretch}{1.3}
\begin{ruledtabular}
\begin{tabular}{cr}
$H^{\Gamma^a_8 \Gamma^c_8}_{33}$({\bf k}) & \parbox[t]{2.5in}{\raggedright $
   E_{33} (\oneq) g_0 $ } \\
$H^{\Gamma^a_8 \Gamma^c_8}_{31}$({\bf k}) & \parbox[t]{2.5in}{\raggedright $ i \left[ E_{3\bar{1}} (\oneq) +
   \sqrt{3/2} E_{3\bar{3}} (\oneq) \right] \left( g_x + i g_y \right) $ } \\
$H^{\Gamma^a_8 \Gamma^c_8}_{3\bar{1}}$({\bf k}) & \parbox[t]{2.5in}{\raggedright $ i E_{3\bar{1}} (\oneq) g_z $ } \\
$H^{\Gamma^a_8 \Gamma^c_8}_{3\bar{3}}$({\bf k}) & \parbox[t]{2.5in}{\raggedright $
   i E_{3\bar{3}} (\oneq) \left( g_x - i g_y \right) / \sqrt{2} $ } \\
$H^{\Gamma^a_8 \Gamma^c_8}_{13}$({\bf k}) & \parbox[t]{2.5in}{\raggedright $ -i \left[ E_{3\bar{1}} (\oneq) +
   \sqrt{1/6} E_{3\bar{3}} (\oneq) \right] \left( g_x - i g_y \right) $ } \\
$H^{\Gamma^a_8 \Gamma^c_8}_{11}$({\bf k}) & \parbox[t]{2.5in}{\raggedright $
   E_{33} (\oneq) g_0 $ } \\
$H^{\Gamma^a_8 \Gamma^c_8}_{1\bar{1}}$({\bf k}) & \parbox[t]{2.5in}{\raggedright $
   -i E_{3\bar{3}} (\oneq) \left( g_x + i g_y \right) / \sqrt{2} $ } \\
$H^{\Gamma^a_8 \Gamma^c_8}_{1\bar{3}}$({\bf k}) & \parbox[t]{2.5in}{\raggedright $ i \left[ E_{3\bar{1}} (\oneq) +
   2 \sqrt{2/3} E_{3\bar{3}} (\oneq) \right] g_z $ } \\
\multicolumn{2}{l}{
    $H^{\Gamma^a_8 \Gamma^c_8}_{\bar{i}\bar{j}} (k_x,k_y,k_z) = H^{\Gamma^a_8 \Gamma^c_8}_{ij} (k_x,-k_y,-k_z)$
} \\
$H^{\Gamma^a_7 \Gamma^c_7}_{\uparrow \uparrow}$({\bf k}) & \parbox[t]{2.5in}{\raggedright $
   E_{\uparrow \uparrow} (\oneq) g_0 $ } \\
$H^{\Gamma^a_7 \Gamma^c_7}_{\uparrow \downarrow}$({\bf k}) & \parbox[t]{2.5in}{\raggedright $ 0 $ } \\
\multicolumn{2}{l}{
    $H^{\Gamma^a_7 \Gamma^c_7}_{\downarrow \downarrow} (k_x,k_y,k_z) =
    H^{\Gamma^a_7 \Gamma^c_7}_{\uparrow \uparrow} (k_x,-k_y,-k_z)$;
} \\
\multicolumn{2}{l}{
    $H^{\Gamma^a_7 \Gamma^c_7}_{\downarrow \uparrow} (k_x,k_y,k_z) =
    H^{\Gamma^a_7 \Gamma^c_7}_{\uparrow \downarrow} (k_x,-k_y,-k_z)$
} \\
$H^{\Gamma^a_8 \Gamma^c_7}_{3 \uparrow}$({\bf k}) & \parbox[t]{2.5in}{\raggedright $
   -i E_{3 \uparrow} (\oneq) \left( g_x + i g_y \right) / \sqrt{2} $ } \\
$H^{\Gamma^a_8 \Gamma^c_7}_{1 \uparrow}$({\bf k}) & \parbox[t]{2.5in}{\raggedright $ 0 $ } \\
$H^{\Gamma^a_8 \Gamma^c_7}_{\bar{1} \uparrow}$({\bf k}) & \parbox[t]{2.5in}{\raggedright $
   -i \sqrt{3/2} E_{3 \uparrow} (\oneq) \left( g_x - i g_y \right) $ } \\
$H^{\Gamma^a_8 \Gamma^c_7}_{\bar{3} \uparrow}$({\bf k}) & \parbox[t]{2.5in}{\raggedright $
   -i \sqrt{2} E_{3 \uparrow} (\oneq) g_z $ } \\
\multicolumn{2}{l}{
    $H^{\Gamma^a_8 \Gamma^c_7}_{\bar{i}\downarrow} (k_x,k_y,k_z) =
    -H^{\Gamma^a_8 \Gamma^c_7}_{i\uparrow} (k_x,-k_y,-k_z)$
} \\
$H^{\Gamma^a_7 \Gamma^c_8}_{\uparrow 3}$({\bf k}) & \parbox[t]{2.5in}{\raggedright $
   i E_{\uparrow 3} (\oneq) \left( g_x - i g_y \right) / \sqrt{2} $ } \\
$H^{\Gamma^a_7 \Gamma^c_8}_{\uparrow 1}$({\bf k}) & \parbox[t]{2.5in}{\raggedright $ 0 $ } \\
$H^{\Gamma^a_7 \Gamma^c_8}_{\uparrow \bar{1}}$({\bf k}) & \parbox[t]{2.5in}{\raggedright $
   i \sqrt{3/2} E_{\uparrow 3} (\oneq) \left( g_x + i g_y \right) $ } \\
$H^{\Gamma^a_7 \Gamma^c_8}_{\uparrow \bar{3}}$({\bf k}) & \parbox[t]{2.5in}{\raggedright $
   i \sqrt{2} E_{\uparrow 3} (\oneq) g_z $ } \\
\multicolumn{2}{l}{
    $H^{\Gamma^a_7 \Gamma^c_8}_{\downarrow\bar{i}} (k_x,k_y,k_z) =
    -H^{\Gamma^a_7 \Gamma^c_8}_{\uparrow i} (k_x,-k_y,-k_z)$
} \\
\end{tabular}
\end{ruledtabular}
\caption{Matrix elements for the zinc blende structure (double group) for quasi-atomic orbitals. The remaining matrix elements can be obtained from the hermiticity of the Hamiltonian in the Bloch sum representation. The definitions $g_0 = 4 ( \cos \xi \cos \eta \cos \zeta - i \sin \xi \sin \eta \sin \zeta )$, $g_x = 4 (- \cos \xi \sin \eta \sin \zeta + i \sin \xi \cos \eta \cos \zeta )$ (and cyclic permutations), are made, where $\xi \equiv k_x a/4$, $\eta \equiv k_y a/4$ and $\zeta \equiv k_z a/4$.}
\label{tab:Td_double_QAO}
\end{table}

\subsection{Effective bond orbitals and \kp}
\label{ssec:EBOkp}

We can obtain further insight into the nature of the L\"owdin orbitals (LOs) used as a basis set by noting that, for orbitals associated with the primitive cells [effective bond orbitals (EBOs)] and at ${\bf k}=0$, we have
\begin{multline}
H^{\mu\mu'}_{ii'} ({\bf k}=0) = \\
    \sum_{{\bf r}_j} \frac{1}{N({\bf r}_j)}
    \sum_{G} D(G)^{\mu}_{is}  E^{\mu\mu'}_{ss'} ({\bf r}_j)  D(G^{-1})^{\mu'}_{s'i'} ,
\end{multline}
and by use of the orthogonality theorem for the irreducible representations~\cite{Tung} we arrive at
\begin{equation}
H^{\mu\mu'}_{ii'} ({\bf k}=0) = \delta_{\mu\mu'} \delta_{ii'}
    \sum_{{\bf r}_j ,s} \frac{N(\mathcal{G}_0)}{N({\bf r}_j) d^\mu} E^{\mu\mu'}_{ss} ({\bf r}_j) ,
\end{equation}
where $N(\mathcal{G}_0)$ is the number of operations of the point group and $d^\mu$ is the dimensionality of the $\mu$-th irrep.

Thus, if one is constructing a model where there are no repeated irreps, we see that the Bloch sums of EBOs will exactly diagonalize the Hamiltonian at ${\bf k}=0$. In other words, the Bloch sums of EBOs will be the zone center solutions, making the connection between the EBOs and the \kp\ basis evident.


\section{Summary}

In conclusion, we have constructed full-zone spin Hamiltonians for [001], [110] and [111] zinc blende quantum wells. We then performed small ${\bf k}$ expansions of those Hamiltonians about the zone center, yielding their \kp\ counterparts. The \kp\ Hamiltonians thus obtained present spin-dependent terms that had not been previously described in the literature. In particular, we see that the Rashba Hamiltonian can be supplemented with third order terms, which will need to be included in calculations where other sources of spin splitting are considered up to that order. We also generate additional, growth direction-dependent $\mathcal{O} (k^3)$ contributions to the Dresselhaus Hamiltonian. The method we have employed is not restricted to these particular cases, as it extends the tight binding formalism to include the treatment of spin by using double group representations. This guarantees the systematic inclusion of all spin-related effects in the bands. Thus, our work can serve as the basis for numerical studies of large scale nanostructures where spin effects are important, as well as an analytic tool for predicting spin properties in reduced-symmetry systems.

\acknowledgments
This work was supported in part by the U.S. Department of Energy under Contract No.\ DE-AC03-76SF00098, by the Defense Advanced Research Projects Agency (DARPA) under Contract No.\ DAAD19-01-1-0324, through HRL Laboratories by DARPA under Contract No.\ MDA972-01-C-0002 and by the European Commission's Marie Curie International Reintegration Grant No.\ MIRG-CT-2005-017198. XC is a Spain's Ministry of Education and Science Ram\'on y Cajal fellow. A part of this work was carried out at the Jet Propulsion Laboratory, California Institute of Technology, through an agreement with the National Aeronautics and Space Administration.


\appendix

\section{Empirical tight-binding review}
\label{ssec:ETB_review}

We review the empirical tight-binding method in order to set forth the notation employed in the paper. Following Ref.~\onlinecite{SlaterKoster1954}, we choose as our basis a collection of L\"owdin-symmetrized orbitals~\cite{Lowdin1950} $\ket{ \phi^{\mu}_i ; {\bf R}_j }$, denoting an orbital centered on the lattice site ${\bf R}_j$ that transforms as the $i$-th state of a basis set for the $\mu$-th irrep of the point group symmetry of the crystal ($\mu$ and $i$ jointly define the band index of the state). Explicitly, they are given by
\begin{equation}
\ket{ \phi^{\mu}_i ; {\bf R}_j } =
   \ket{ \varphi^{\mu'}_{i'} ; {\bf R}_{j'} } \left( S^{-1/2} \right)^{\mu'\mu}_{i'i,{\bf R}_{j'} {\bf R}_j } ,
\label{eq:Bloch_sum}
\end{equation}
where $S^{-1/2}$ is the inverse of the square root of the overlap matrix $S$, with elements $S^{\mu'\mu}_{i'i,{\bf R}_{j'} {\bf R}_j } = \braket{ \varphi^{\mu'}_{i'} ; {\bf R}_{j'} }{ \varphi^{\mu}_{i} ; {\bf R}_{j} }$, and $\ket{ \varphi^{\mu}_{i} ; {\bf R}_{j} }$ is the atomic orbital (or, in general, a state with finite overlap, {\it e.g.} a bond orbital) analogous to $\ket{ \phi^{\mu}_i ; {\bf R}_j }$. This is well defined, as the L\"owdin symmetrization procedure ensures that the state will have the same symmetry properties as the underlying atomic orbital from which it is constructed~\cite{SlaterKoster1954}. As shown by Chang~\cite{Chang1988} and exploited in Sec.~\ref{sec:application}, under certain conditions L\"owdin orbitals (LOs) provide the simplest connection to the \kp\ method. Another useful property is that LOs centered at different sites are orthogonal.

For the case of a bulk material, SK proceeded to construct bulk Bloch sums from the LOs
\begin{equation}
\ket{ \phi^{\mu}_{i,{\bf k}} } = \frac{1}{\sqrt{N}} \sum_{{\bf R}_j} e^{i {\bf k} \cdot {\bf R}_j}
     \ket{ \phi^{\mu}_i ; {\bf R}_j },
\end{equation}
where $N$ is the number of lattice sites in the crystal and ${\bf k}$ lies within the first Brillouin zone. When the Schr\"odinger equation $H \ket{\psi_{\bf k}} = E \ket{\psi_{\bf k}}$ is expanded in terms of Bloch sums, it becomes an eigenvalue problem
\begin{equation}
\sum_{\mu',i'} \melem{ \phi^{\mu}_{i,{\bf k}} }{ H }{ \phi^{\mu'}_{i',{\bf k}} }
   c^{\mu'}_{i'} ({\bf k}) = E^{\mu}_{i,{\bf k}} c^{\mu}_{i} ({\bf k}) ,
\label{eq:eval}
\end{equation}
where $E^{\mu}_{i,{\bf k}}$ is the energy at point ${\bf k}$ and we have defined a expansion coefficient $c^{\mu}_{i} ({\bf k}) \equiv \braket{ \phi^{\mu}_{i,{\bf k}} }{ \psi_{\bf k} }$. The tight-binding matrix elements are easily seen to be given by
\begin{multline}
H^{\mu\mu'}_{ii'} ({\bf k}) \equiv \melem{ \phi^{\mu}_{i,{\bf k}} }{ H }{ \phi^{\mu'}_{i',{\bf k}} } = \\
    \sum_{{\bf R}_j} e^{i {\bf k} \cdot {\bf R}_j}
    \melem{ \phi^{\mu}_i ; {\bf 0} }{ H }{ \phi^{\mu'}_{i'} ; {\bf R}_j }.
\label{eq:phi_k_melem}
\end{multline}

In empirical tight-binding (ETB), the matrix elements on the right hand side of \refeq{eq:phi_k_melem} are taken as adjustable parameters
\begin{equation}
E^{\mu\mu'}_{ii'} ({\bf R}_j) \equiv \melem{ \phi^{\mu}_i ; {\bf 0} }{ H }{ \phi^{\mu'}_{i'} ; {\bf R}_j } .
\label{eq:ETB_pars}
\end{equation}

The sum over the neighboring sites in \refeq{eq:phi_k_melem} can be rewritten as
\begin{equation}
\melem{ \phi^{\mu}_{i,{\bf k}} }{ H }{ \phi^{\mu'}_{i',{\bf k}} } =
    \sum_{{\bf r}_j} \frac{1}{N({\bf r}_j)} \sum_{G} e^{i {\bf k} \cdot G {\bf r}_j}
    E^{\mu\mu'}_{ii'} (G {\bf r}_j)
\label{eq:phi_k_melem2}
\end{equation}
where ${\bf r}_j$ is the collection of Bravais lattice sites not related by point group ($\mathcal{G}_0$) operations $G$ (\ie\ sum over lattice points belonging to different stars) and $N({\bf r}_j)$ is the number of operations that leave ${\bf r}_j$ invariant.

Since $G$ acting on a state $\ket{ \phi^{\mu'}_{i'} ; {\bf r}_j }$ changes both the type of state and its lattice point, it is easy to see that $T_{G{\bf r}_j} = G T_{{\bf r}_j} G^{-1}$, where $T_{{\bf r}_j}$ carries a state $\ket{ \phi^{\mu'}_{i'} ; {\bf 0} }$ into $\ket{ \phi^{\mu'}_{i'} ; {\bf r}_j }$. Thus, we have
\begin{multline}
E^{\mu\mu'}_{ii'} (G {\bf r}_j) =
    \melem{ \phi^{\mu}_i ; {\bf 0} }{ H T_{G{\bf r}_j} }{ \phi^{\mu'}_{i'} ; {\bf 0} } = \\
    \melem{ \phi^{\mu}_i ; {\bf 0} }{ H G T_{{\bf r}_j} G^{-1} }{ \phi^{\mu'}_{i'} ; {\bf 0} } .
\end{multline}
Since $G \in \mathcal{G}_0$, we have that $H G = G H$, therefore we arrive at
\begin{equation}
E^{\mu\mu'}_{ii'} (G {\bf r}_j) = %
    D(G)^{\mu}_{is}  E^{\mu\mu'}_{ss'} ({\bf r}_j)  D(G^{-1})^{\mu'}_{s'i'} ,
\label{eq:Eirrep}
\end{equation}
where we make use of
\begin{equation}
G \ket{ \phi^{\mu}_{i} ; {\bf r}_j } = \ket{ \phi^{\mu}_{s} ; G {\bf r}_j } D(G)^{\mu}_{si}
\label{eq:transf_LO}
\end{equation}
and $D(G)^{\mu}_{si}$ is the element for the matrix corresponding to operation $G$ of the $\mu$-th irrep of the crystal point group. We substitute \refeq{eq:Eirrep} into \refeq{eq:phi_k_melem2} to obtain
\begin{multline}
H^{\mu\mu'}_{ii'} ({\bf k}) = \\
 \sum_{{\bf r}_j} \frac{1}{N({\bf r}_j)} \sum_{G} e^{i k_t D(G)^{\rm vec}_{tt'} r_{j,t'}}
    D(G)^{\mu}_{is}  E^{\mu\mu'}_{ss'} ({\bf r}_j)  D(G^{-1})^{\mu'}_{s'i'} ,
\label{eq:dirrep}
\end{multline}
where $D(G)^{\rm vec}$ is the representation of $G$ for polar vectors, which in general will not be an irrep of the point group. This form allows for an easier computer implementation since there is no need to keep track of which neighbor positions have been visited and which not, and a single loop over all symmetry operations can be used.

SK present in their tables results for the Hamiltonian matrix elements computed with a method analogous to (\ref{eq:dirrep}), but only single group irreps are used, which effectively does away with spin. In Sec.~\ref{sec:application} we carry out calculations with double group irreps, thus taking into account all spin effects within the single particle approximation.

Note that this procedure can deal with general kinds of crystal lattices, not only those with a single atom per primitive cell, because any symmetry reduction due to the atom basis attached to each lattice site is mimicked by the imposed symmetry of the L\"owdin orbitals. For example, in a zinc blende structure orbitals transforming according to $T_d$ (i.e. without definite parity) are attached to {\em fcc} lattice sites.

The inclusion of more than one type of orbital per primitive cell (\ie\ need to have orbitals attached to atomic sites rather than lattice sites) is easily achieved by introducing and extra index $\alpha$ for the atom type in the states and having ${\bf R}_j \rightarrow {\bf R}_j + {\bf d}_{\alpha'} - {\bf d}_\alpha$ in \refeq{eq:phi_k_melem}, where ${\bf d}_\alpha$ is the offset of species $\alpha$ with respect to the lattice site. After that the derivation until Eqs.~(\ref{eq:Eirrep}) and (\ref{eq:dirrep}) continues unchanged.

\section{Symmetry considerations}
\label{ssec:symm_cons}

It is useful to use all the available symmetries to reduce as much as possible the number of independent $E^{\mu\mu'}_{ss'} ({\bf r}_j)$'s. This becomes specially important when ETB is applied to lower dimensionality structures, as then the sum (\ref{eq:phi_k_melem}) is not carried out for some directions (\ie\ planar, linear or point orbitals are constructed).

\subsubsection{Point group symmetry}

When a $\mathcal{G}_0$ operation $G$ is such that $G{\bf r}_j = {\bf r}_j$ for a given ${\bf r}_j$, we obtain a number of consistency requirements for the neighbor matrix elements $E^{\mu\mu'}_{ii'} ({\bf r}_j)$ from \refeq{eq:Eirrep}:
\begin{equation}
E^{\mu\mu'}_{ii'} ({\bf r}_j) =
    D(G)^{\mu}_{is}  E^{\mu\mu'}_{ss'} ({\bf r}_j)  D(G^{-1})^{\mu'}_{s'i'} .
\label{eq:G_sym}
\end{equation}

%

\subsubsection{Time reversal}

For double group irreps, the basis states can be labeled~\cite{KosterDimmockWheeler1963} by $i=-q_\mu,\ldots,-1/2,1/2,\ldots,q_\mu$. For example, the top of the valence band in zinc blendes would have $q_{\mu} = 3/2$, the bottom of the conduction band $q_{\mu} = 1/2$, etc. We employ the phase convention~\cite{KosterDimmockWheeler1963}
\begin{equation}
\Theta \ket{ \phi^{\mu}_i ; {\bf r}_j } \equiv \ket{ \tilde{\phi}^{\mu}_i ; {\bf r}_j } =
    (-1)^{q_\mu - i} \ket{ \phi^{\mu}_{-i} ; {\bf r}_j }
\end{equation}
for the action of the time reversal operator $\Theta$. Since $\Theta$ is antiunitary~\cite{Sakurai}, we will have
\begin{equation}
E^{\mu\mu'}_{ii'} ({\bf r}_j) = (-1)^{q_\mu + q_{\mu'}} (-1)^{i + i'} {E^{\mu \mu'}_{-i,-i'}}^\ast ({\bf r}_j) .
\label{eq:TR_sym}
\end{equation}

\subsubsection{Rotation followed by translation}

If there is an operation such that $G{\bf r}_j = -{\bf r}_j$, we can combine \refeq{eq:Eirrep} with the translational symmetry requirement $E^{\mu\mu'}_{ii'} (-{\bf r}_j) = {E^{\mu'\mu}_{i'i}}^\ast ({\bf r}_j)$ to obtain
\begin{equation}
{E^{\mu'\mu}_{i'i}}^\ast ({\bf r}_j) = D(G)^{\mu}_{is}  E^{\mu\mu'}_{ss'} ({\bf r}_j)  D(G^{-1})^{\mu'}_{s'i'} .
\label{eq:rotot_sym}
\end{equation}
Now we can remove the complex conjugation operation by employing time reversal, \refeq{eq:TR_sym}. The result is
\begin{multline}
E^{\mu'\mu}_{-i',-i} ({\bf r}_j) = \\
    (-1)^{q_\mu + q_{\mu'}} (-1)^{i + i'} D(G)^{\mu}_{is}  E^{\mu\mu'}_{ss'} ({\bf r}_j)  D(G^{-1})^{\mu'}_{s'i'} .
\label{eq:rototTR_sym}
\end{multline}

If the crystal has an inversion center, we will always be able to pick $G=I$, where $I$ is the inversion operator. Then, since every irrep $\mu$ of a crystal with inversion will have a definite parity $\pi_\mu$, we arrive at
\begin{equation}
E^{\mu'\mu}_{-i',-i} ({\bf r}_j) =
    (-1)^{q_\mu + q_{\mu'}} (-1)^{i + i'} \pi_{\mu} \pi_{\mu'} E^{\mu\mu'}_{ii'} ({\bf r}_j) .
\label{eq:I_sym}
\end{equation}

\subsubsection{Bloch sums}

It is also convenient to state explicitly the transformation properties of Bloch sums and their matrix elements. This can reduce the number of elements that need to be specified. Combining Eqs.~(\ref{eq:Bloch_sum}) and (\ref{eq:transf_LO}) it is easy to see that
\begin{equation}
G \ket{ \phi^{\mu}_{i,{\bf k}} } = \ket{ \phi^{\mu}_{s,G {\bf k}} } D(G)^{\mu}_{si} ,
\label{eq:transf_BS}
\end{equation}
which leads to the relationship~\cite{BirPikus}
\begin{equation}
H^{\mu\mu'}_{ii'} ({\bf k}) =
    D(G)^{\mu}_{is}  H^{\mu\mu'}_{ss'} (G^{-1} {\bf k})  D(G^{-1})^{\mu'}_{s'i'} ,
\label{eq:sym_Hmelem}
\end{equation}
constraining the values of the Hamiltonian matrix elements between Bloch sums.

\section{Models with a finite number of bands}

The procedure described in Appendix~\ref{ssec:ETB_review} will only yield the exact Hamiltonian when an infinite number of L\"owdin orbitals (LOs) is included and interaction with all neighbors is accounted for. Of course, the localized character of the LOs will limit the neighbor distance at which there is non-negligible overlap. On the other hand, since the number of included neighbors is directly related to the number of Fourier components of the bands, the effect of coupling to bands outside the model---which will add to the number of Fourier components---can be modeled by increasing the range of the effective Hamiltonian interaction~\cite{TingChang1987}.

In the extreme case where only two bands are included (say, spin up and spin down), a large number of included neighbors will provide a good description of the bands. In this context Kramers degeneracy is easily shown for doubly degenerate irreps (\ie\ doubly degenerate at the zone center, but not {\it a priori} at a general ${\bf k}$ point). Labeling $\mu \rightarrow 2 q_\mu$, time reversal combined with inversion symmetry [\refeq{eq:I_sym}]  implies $E^{11}_{\uparrow \uparrow} ({\bf r}_j) = E^{11}_{\downarrow \downarrow} ({\bf r}_j)$ and $E^{11}_{\uparrow \downarrow} ({\bf r}_j) = E^{11}_{\downarrow \uparrow} ({\bf r}_j) = 0$. Thus, the spin $2 \times 2$ Hamiltonian takes on the form
\begin{equation}
H^{11}_{\text{inv}} =
\begin{pmatrix}
H_{\uparrow \uparrow} ({\bf k}) & 0 \\
0 & H_{\uparrow \uparrow} ({\bf k})
\end{pmatrix} ,
\end{equation}
which shows that the bands will indeed be degenerate at a general ${\bf k}$ point.


\begin{thebibliography}{45}
\bibitem[{\citenamefont{de~Andrada~e Silvus}(1992)}]{endnote45}{The
derivation of the last term in \refeq{eq:D2d_kp} from bulk properties
requires constructing the bulk spin Hamiltonian to $\mathcal{O} (k^5)$
and only then replacing $\hat{k}_z^{2n} \rightarrow \left\langle
\hat{k}_z^{2n} \right\rangle $.}
\end{thebibliography}
\end{document}